\documentclass[twocolumn]{IEEEtran}
\usepackage{amsmath}
\usepackage{amsthm}
\usepackage{amsfonts}
\usepackage{amssymb}
\usepackage{mathrsfs}
\usepackage{cases}
\usepackage{cite}
\usepackage{graphicx}
\usepackage{mathrsfs}
\usepackage{subfigure}
\usepackage{epstopdf}
\usepackage{color}
\usepackage{bm}
\usepackage{caption}
\usepackage{multirow}
\usepackage{diagbox}
\allowdisplaybreaks[4]
\theoremstyle{remark}

\begin{document}

\title{\LARGE Passive Beamforming for IRS Aided Wireless Networks
\author{Ke-Wen Huang \hspace{0.05in}and  Hui-Ming Wang,~\IEEEmembership{Senior Member,~IEEE}
}
\thanks{
The authors are with the School of Information and Communication Engineering, and also with the Ministry
of Education Key Lab for Intelligent Networks and Network Security, Xi'an Jiaotong University, Xi'an 710049, Shaanxi, P. R. China
(e-mail:
xjtu-huangkw@outlook.com, xjbswhm@gmail.com).
}
}
\maketitle

\begin{abstract}
In this letter, we design passive beamforming in an intelligent reflecting surface (IRS) assisted multiple-user wireless network.
Two different scenarios are considered, namely, multicasting and multi-user downlink transmission.
We optimize the passive beamforming vector of the IRS to maximize the smallest signal-to-noise ratio of the users in both scenarios. Based on the \emph{alternating direction method of multipliers} algorithm, a low complexity method is designed to iteratively solve the established problem.
In each iteration of the proposed method, the solution is in closed form, and thus the computation complexity is low. Numerical results are presented to show the efficiency of the proposed method.
\end{abstract}


\section{Introduction}
Recently, intelligent reflecting surface (IRS), improving the wireless environment by reflecting incident
electromagnetic waves in a controllable manner, has gained considerable research attention \cite{C.Liaskos2018,Q.WuCMCM2020}.
Intuitively, if the reflecting coefficients (RCs) of the IRS, namely, passive beamforming, are properly designed, then
the signal reflected by the IRS can be coherently superimposed on the signal from other propagation paths at the intended receiver, which significantly enhances the signal strength.

To make full use of the IRS, some recent efforts have been devoted to design and optimization for IRS-aided wireless communications  \cite{Q.WuTWC2019,Y.ZhengWCL,
C.HuangTWC2019,X.GuanCL2020,LimengDongFx}.
In \cite{Q.WuTWC2019}, transmissions from a multiple-antenna transmitter to multiple receivers were considered. The authors minimized the transmit power by optimizing the transmit beamforming at the transmitter and the passive beamforming at the IRS.
%The transmit power minimization problem studied in \cite{Q.WuTWC2019} was futher extended in \cite{Q.WuICASSP2019} to a more practical situation where the phase shifts of the IRS can only be chosen from a pre-given discrete set due to the hardware limitation.
Theoretically, \cite{Q.WuTWC2019} showed that with the help of an IRS, the transmit power can be reduced by a factor of $\frac{1}{N^2}$, where $N$ is the number of the reflecting elements.
In \cite{C.HuangTWC2019}, energy efficiency maximization problem in an IRS-aided wireless system was studied,
and the authors alternatingly optimized the power allocation at the transmitter and the passive beamforming at the IRS. The numeric results in \cite{C.HuangTWC2019} reveal that with the aided of an IRS, the system achieves higher energy efficiency compared to the case where the IRS is replaced by an amplify-and-forward relay.
In \cite{Y.ZhengWCL}, passive beamforming is designed in wireless powered communication networks, wherein the IRS not only improves the quality of the information-carrying signal but also helps the receivers harvesting wireless energy.
In \cite{X.GuanCL2020}, an IRS-aided cognitive radio communication system was studied, and it was shown that using passive beamforming, the communication performance of a secondary user can be significantly improved.
In \cite{LimengDongFx},  the IRS is used to improve the security of wireless systems.
%In addition to enhancing wireless reliability, IRS has also been considered to safeguard the privacy and security of wireless transmission, see e.g., \cite{Z.Chu2020WCL,H.ShenCL2019}.

Though IRS has been applied to many different scenarios (as introduced above), the method to obtain a good passive beamforming vector has not been well studied. In general, optimizing the passive beamforming involves solving an non-convex problem, which is usually hard to handle.
Existing works usually tackle the non-convexity by using the technique of semidefine relaxation (SDR) and transforming the optimization problem into a convex semidefine programming (SDP), see e.g., \cite{Q.WuTWC2019,Y.ZhengWCL}. However, in general, the computational complexity of the SDR-based method is very high. Specifically, using the SDR-based method, $O(N^2)$ variables need to be optimized, and thus does not apply to the case with massive reflecting elements.
Besides, the SDR-based method may not generate rank-one solution, meaning that techniques, such like Gaussian randomization, is required to recover a feasible solution, which further increases the computational burden.
There are indeed some low complexity designs \cite{C.HuangTWC2019,X.GuanCL2020}, which are, however, restricted to some special scenarios,
for example, when direct links are absent \cite{C.HuangTWC2019} and when only one secondary user are considered \cite{X.GuanCL2020}.

Motivated by the observations above, in this letter, we aim at designing low complexity method to obtain good passive beamforming vector for IRS-aided multiple-user wireless networks. Two different communication scenarios, namely, multicasting and multi-user downlink transmission, are considered. For both cases, we maximize the smallest signal-to-noise ratio (SNR) of the users, and low complexity methods are designed based on the \emph{alternating direction method of multipliers} (ADMM) algorithm \cite{S.boydADMM2011}.
In the following, we first introduce the system model, and then discuss the proposed methods.

\section{System Model}
We consider that a base station (BS), equipped with $M$ antennas, transmits to $K$ single-antenna users under the help of an IRS that consists of $N$ reflecting elements.
Denote by $\bm{f}_k\in\mathcal{C}^{M\times1}$, $\bm{h}_k\in\mathcal{C}^{N\times1}$,  $\bm{G}\in\mathcal{C}^{N \times M}$  the channels from the BS to the $k$-th user, from the IRS to the $k$-th user, and  from the BS to the IRS, respectively. The channel are assumed to be static and are known to the BS.
%A comprehensive system model is illustrated in Fig. \ref{Fig:Model}.

We consider two communication scenarios, i.e., multicasting and multi-user downlink transmission. In the case of multicasting, the BS transmits a common message to all users. Denote by $\bm{b}$ the beamforming vector of the BS. The signal-to-noise ratio (SNR) of the $k$-th user is
\begin{align}
\label{SNRMulticast}
{\rm SNR}_k^{(\mathrm{m})} =  |\left(\bm{h}_k^H \bm{\Phi}^H \bm{G} + \bm{f}_k^H\right)\bm{b}|^2/\sigma_k^2,
\end{align}
where $\sigma_k^2$ is the noise power of the $k$-th user and $\bm{\Phi} = \mathrm{diag}(\bm{\phi})$ consists of the RCs of the IRS with $\bm{\phi} = [\phi_1,\phi_2,\cdots,\phi_N]^T$ and $|\phi_n| \leq 1$ for $1\leq n\leq N$.
In the case of multi-user downlink transmission, the BS transmits different messages to different users. Denote by $\bm{b}_k$ the beamforming vector for the $k$-th user. Treating the interference as noise, the SNR of the $k$-th user is
\begin{align}
\label{SNRbroadcast}
{\rm SNR}_k^{(\mathrm{d})} = \frac{|\left(\bm{h}_k^H \bm{\Phi}^H \bm{G} + \bm{f}_k^H\right)\bm{b}_k|^2}{\sigma_k^2 + \sum_{k^{\prime} \neq k} |\left(\bm{h}_k^H \bm{\Phi}^H \bm{G} + \bm{f}_k^H\right)\bm{b}_{k^{\prime}}|^2}.
\end{align}

In this letter, we aim to optimize the passive beamforming vector $\bm{\phi}$ to improve the communication performance. Note that in practice, the transmit beam of the BS
and the passive beam of the IRS can be jointly optimized.
However, in this letter, we only focus on the sub-problem of optimizing $\bm{\phi}$. This is because such a joint optimization problem is usually treated in an alternating manner, and when $\bm{\phi}$ is fixed, the problem degrades to conventional transmit beamforming problem which has been extensively investigated. In this letter, we focus on the following max-min problem,
\begin{subequations}
\label{FirstForm}
\begin{align}
\label{OverallObj}
\max_{\bm{\phi}}&\quad\quad\min_{k}~ {\rm SNR}_k^{(\mathrm{s})}(\bm{\phi}), \\
{\rm s.t.}& \quad\quad|\phi_n|^2 \leq 1,\quad \text{for }\forall n, \label{IdealRS}
\end{align}
\end{subequations}
where $\mathrm{s} = \mathrm{m}$ stands for the multicasting scenario, and $\mathrm{s} = \mathrm{d}$ stands for the multi-user downlink transmission scenario.
In this letter, $\bm{b}$ and $\{\bm{b}_k\}_{k=1}^K$ are assumed to be fixed, and thus the formulated optimization problem \eqref{FirstForm} is irrespective of the active beamforming scheme adopted by the BS.

By \eqref{IdealRS}, we consider ideal IRS with continuous RCs.
We note that for implementation simplicity, some literature assumed that the amplitude of the RCs are fixed as $1$, meaning that the RCs are subject to the following constraints
\begin{align}
|\phi_n| = 1,\quad \text{for }\forall n. \label{NotIdealIRS}
\end{align}
However, \eqref{NotIdealIRS} is not a convex constraint and is generally hard to handle directly.
In this letter, we focus on optimizing $\bm{\phi}$ under constraint \eqref{IdealRS}. Though we do not directly handle constraint \eqref{NotIdealIRS}, we will check the communication performance under constraint \eqref{NotIdealIRS} by directly projecting the solution to \eqref{FirstForm} into the set of $\bm{\phi}$ defined by \eqref{NotIdealIRS}.
In the following two sections, we present our method to solve \eqref{FirstForm} in the multicasting and the multi-user downlink transmission scenarios, respectively.

\emph{Remark:} Problem \eqref{FirstForm} differs form the SNR balancing problem in \cite{A.WieselTSP} and cannot be solved by the method in \cite{A.WieselTSP}.
Specifically, in \eqref{FirstForm}, a single passive beam is designed for all users, while in \cite{A.WieselTSP}, different users are associated with different beamforming vectors. As a result, \eqref{FirstForm} can not be transformed into a generalized eigenvalue problem as in \cite{A.WieselTSP}.

%\begin{figure}[!t]
%\begin{center}
%\includegraphics[width=2 in]{Model.eps}
%\caption{System model, where an access point communicates with multiple users with the aided of an IRS.}\label{Fig:Model}
%\end{center}
%\vspace{-5mm}
%\end{figure}

\section{Multicasting}
In this section, we present a low complexity method for optimizing the passive beamforming vector. Before presenting our method, we first introduce the widely used SDR-based method, which will be used as a benchmark for comparison.
\subsection{Benchmark SDR-based method}
We reformulate $\mathrm{SNR}_k^{(\rm m)}$ as,
\begin{align}
\mathrm{SNR}_k^{(\rm m)} = | \bm{\alpha}_k^H \bm{\phi} + \beta_k|^2 = \bar{\bm{\phi}}^H\bar{\bm{\alpha}}_k\bar{\bm{\alpha}}_k^H\bar{\bm{\phi}}
= \mathrm{Tr}\left(\bm{\Psi} \bm{\Xi}_k \right),
\end{align}
where $\bm{\alpha}_k \triangleq \frac{1}{\sigma_k}\bm{H}_k^H\bm{G}\bm{b} $, $\bm{H}_k \triangleq \mathrm{diag}\{\bm{h}_k\}$, $\beta_k \triangleq \frac{1}{\sigma_k}\bm{b}^H\bm{f}_k$, $\bar{\bm{\phi}} \triangleq [\bm{\phi}^T,1]^T$, $\bar{\bm{\alpha}}_k \triangleq [\bm{\alpha}_k^T,\beta_k^*]^T$,
$\bm{\Psi} \triangleq \bar{\bm{\phi}}\bar{\bm{\phi}}^H$, and $\bm{\Xi}_k \triangleq \bar{\bm{\alpha}}_k\bar{\bm{\alpha}}_k^H$. Then, by using the technique of SDR, i.e., namely, neglecting the rank-one constraint on $\bm{\Psi}$, \eqref{FirstForm} becomes
\begin{subequations}
\label{SDPBenchmark}
\begin{align}
\max\limits_{\bm{\Psi}\succeq \bm{0};\gamma} \quad\gamma, \quad {\rm s.t.} \ &\quad  \mathrm{Tr}\left(\bm{\Psi} \bm{\Xi}_k \right) \geq \gamma,\quad \text{for }\forall k,\label{SNRconstraint}\\
&\quad  [\bm{\Psi}]_{n,n}\leq 1,\quad \text{for }\forall 1\leq n\leq N,\label{SNRconstraintIneqaulity}\\
&\quad  [\bm{\Psi}]_{N+1,N+1} =  1.
\end{align}
\end{subequations}
This is a standard convex SDP, and can be solved by mathematic tool such as CVX \cite{M.GrantCVX}.
Note that if constraint \eqref{NotIdealIRS} is considered in replacement of \eqref{IdealRS}, the resulting optimization problem can still be transformed into an SDP in the form of \eqref{SDPBenchmark} except that the inequality constraint \eqref{SNRconstraintIneqaulity} is replaced by $[\bm{\Psi}]_{n,n} = 1$ for $\forall 1\leq n\leq N$.
We point out that such an SDP is actually equivalent to \eqref{SDPBenchmark} in viewing the fact that the left-hand-side of \eqref{SNRconstraint} is increasing with $[\bm{\Psi}]_{n,n}$ for $\forall n$, and thus all the inequality constraints in \eqref{SNRconstraintIneqaulity} are active at the optimal point.
Note that due to the neglected rank-one constraint on $\bm{\Psi}$, \eqref{SDPBenchmark} is not equivalent to \eqref{FirstForm}, and if the optimal solution to \eqref{SDPBenchmark} is not rank-one, then technique such as Gaussian randomization is required to generate a rank-one solution.

\subsection{The proposed ADMM-based method}
By introducing a slack variable $\gamma$, \eqref{FirstForm} is  equivalent to,
\begin{align}
\label{FirstFormReForm}
\max_{\bm{\phi},\gamma}~\gamma,\quad
{\rm s.t.}~\left\{
\begin{aligned}
& {\rm SNR}_k^{(\mathrm{m})}(\bm{\phi}) \geq \gamma, \\
& |\phi_n|^2 \leq 1,\quad \text{for }\forall n.
\end{aligned}\right.
\end{align}
Now, we focus on problem \eqref{FirstFormReForm}.
In fact, the difficulty of solving \eqref{FirstFormReForm} lies in the non-convexity of ${\rm SNR}_k^{(\mathrm{m})}(\bm{\phi})$.
To tackle this problem, we follow the principle of \emph{successive convex approximation (SCA)}, see. e.g., \cite{M.Razaviyayn}.
In brief, the SCA method handles non-convex optimization problem by replacing the non-convex part with some properly selected convex approximations.
In our case, ${\rm SNR}_k^{(\mathrm{m})}(\bm{\phi})$ can be approximated by its the first order Taylor expansion at some feasible point $\bm{\phi}_e$, denoted by $\underline{{\rm SNR}_k^{(\mathrm{m})}}(\bm{\phi},\bm{\phi}_e)$, i.e.,
\begin{align}
{\rm SNR}_k^{(\mathrm{m})}(\bm{\phi}) \geq \underline{{\rm SNR}_k^{(\mathrm{m})}}(\bm{\phi},\bm{\phi}_e) \triangleq2\Re\{ \bm{t}_k^H \bm{\phi} \} + s_k.
\end{align}
where $\bm{t}_k  \triangleq \left( \bm{\alpha}_k^H\bm{\phi}_e
+ \beta_k\right)\bm{\alpha}_k $ and
$s_k \triangleq |\beta_k|^2 - |\bm{\alpha}_k^H \bm{\phi}_e|^2$.
And we obtain the following convex problem,
\begin{align}
\label{MultiCastSCAed}
\max_{\bm{\phi},\gamma}~\gamma,
~{\rm s.t.}\left\{
\begin{aligned}&2\Re\{ \bm{t}_k^H\bm{\phi} \} + s_k \geq \gamma,~ \text{for }\forall k, \\
&|\phi_n|^2 \leq 1,\quad \text{for }\forall n.
\end{aligned}\right.
\end{align}
By repeatedly solving \eqref{MultiCastSCAed}, and setting the point for expansion, i.e., $\bm{\phi}_e$, in each iteration as the optimal solution obtained in the previous iteration, the whole procedure generates a sequence of solution that converge to a Karush-Kuhn-Tucker (KKT) solution of \eqref{FirstForm} \cite[Therorem 1]{M.Razaviyayn}. For more details about the convergence of the SCA method, please refer to \cite{M.Razaviyayn}.

Now, we present an efficient method to solve \eqref{MultiCastSCAed}. The main idea is to use the ADMM algorithm to decompose \eqref{MultiCastSCAed} into multiple parallel sub-problems, each of which is simple and can be solved in closed form.
To do so, we first reformulate \eqref{MultiCastSCAed} as the following equivalent form
\begin{align}
\label{ADMMForMultiCast}
\min_{\mathcal{X},\gamma} ~-\gamma,
~{\rm s.t.}~
\left\{\begin{aligned}
&{\rm C1}: ~g(x_k,z_k) \leq 0,~\text{for }\forall k, \\
&{\rm C2}: ~|y_n|^2 \leq 1,~ \text{for }\forall n, \\
&{\rm C3}: ~\bm{x}  = \bm{T}^H\bm{\phi},~\bm{y} = \bm{\phi},~\bm{z}  = \gamma\bm{1},
\end{aligned}\right.
\end{align}
where $\mathcal{X} \triangleq \{\bm{x},\bm{y},\bm{z} , \bm{\phi}, \gamma\}$, $g(x_k,z_k)\triangleq z_k - 2\Re\{ x_k \} - s_k$, $\bm{T} \triangleq [\bm{t}_1,\bm{t}_2,\cdots,\bm{t}_K]$, and $\bm{1}$ is a column vector with all of its elements being one.
For notational convenience, we use $p_k$ ($p_{j,k}$) to denote the $k$-th element of $\bm{p}$ ($\bm{p}_j$) for any vector $\bm{p}$ ($\bm{p}_j$) and we use $\bm{p}_{\bar{k}}$ ($\bm{p}_{j,\bar{k}}$) to denote the vector obtained by deleting the $k$-th element of $\bm{p}$ ($\bm{p}_j$). By dividing $\mathcal{X}$ in two two groups, i.e., $\mathcal{X}_1 = \{\bm{x},\bm{y},\bm{z}\}$ and $\mathcal{X}_2 = \{\bm{\phi},\gamma\}$, we now use ADMM to solve \eqref{ADMMForMultiCast}.

In principle, the ADMM algorithm solves convex optimization problem by alternatingly updating the primal and dual variables using the Gauss-Seidel method. At the $l$-th iteration, the ADMM algorithm consists of the following steps
\begin{subequations}
\label{ADMMStep}
\begin{align}
&\mathcal{X}_1^{(l+1)} = \mathop{\mathrm{argmin}}\limits_{\mathcal{X}_1}~\mathcal{L}_{\rho}^{(\mathrm{m})} (\mathcal{X}_1,\mathcal{X}_2^{(l)},\bm{u}^{(l)},\bm{v}^{(l)},\bm{w}^{(l)}),\nonumber \\
&\quad\quad\quad\quad\quad\quad  {\rm s.t.}\quad {\rm C1} \text{ and }{\rm C2}\text{ in }\eqref{ADMMForMultiCast},\label{ADMMStep1}\\
&\mathcal{X}_2^{(l+1)} = \mathop{\mathrm{argmin}}\limits_{\mathcal{X}_2}~\mathcal{L}_{\rho}^{(\mathrm{m})} (\mathcal{X}_1^{(l+1)},\mathcal{X}_2,\bm{u}^{(l)},\bm{v}^{(l)},\bm{w}^{(l)}),\label{ADMMStep2}\\
&\bm{u}^{(l+1)} = \bm{u}^{(l)} + \bm{x}^{(l+1)} - \bm{T}^H\bm{\phi}^{(l+1)},\\
&\bm{v}^{(l+1)} = \bm{v}^{(l)} + \bm{y}^{(l+1)} -  \bm{\phi}^{(l+1)},\\
&\bm{w}^{(l+1)} = \bm{w}^{(l)} + \bm{z}^{(l+1)} -  \gamma^{(l)}\bm{1},
\end{align}
\end{subequations}
where $\mathcal{L}_{\rho}^{(\mathrm{m})}$ is the augmented Lagrangian function
\begin{align}
\label{ALForMultiCast}
\mathcal{L}_{\rho}^{(\mathrm{m})} = &-\gamma + \frac{\rho}{2} ||\bm{x}  - \bm{T} ^H \bm{\phi} + \bm{u} ||^2 \nonumber \\
&\quad +\frac{\rho}{2} ||\bm{y} - \bm{\phi} + \bm{v}||^2  +\frac{\rho}{2} ||\bm{z} - \gamma\bm{1} + \bm{w}||^2
\end{align}
with $\rho>0$ being arbitrary and $\{\bm{u},\bm{v},\bm{w}\}$ being the dual variables corresponding to the three equality constraints in \eqref{ADMMForMultiCast}.
Note that the iterations in \eqref{ADMMStep} involves solving two optimization problems, i.e., \eqref{ADMMStep1} and \eqref{ADMMStep2}.
In the following, we show that \eqref{ADMMStep1} and \eqref{ADMMStep2} can be solved in closed form. For notational simplicity, we omit the index of iteration $l$.

 \emph{Solution to \eqref{ADMMStep1}:} with \eqref{ALForMultiCast}, \eqref{ADMMStep1} can be written as
\begin{subequations}
\label{ADMMStep1Solving}
\begin{align}
\min_{\bm{x},\bm{z},\bm{y}} & \quad  \left\|\bm{x}  - \bm{\tau}_1\right\|^2 +  || \bm{z} - \bm{\tau}_2 ||^2 + || \bm{y} - \bm{\tau}_3 ||^2 \\
 {\rm s.t.}& \quad {\rm C1} \text{ and }{\rm C2}\text{ in }\eqref{ADMMForMultiCast}
\end{align}
\end{subequations}
where $\bm{\tau}_1 \triangleq \bm{T}^H\bm{\phi} - \bm{u}$,
$\bm{\tau}_2 \triangleq \gamma\bm{1} - \bm{w}$,
$\bm{\tau}_3 \triangleq \bm{\phi} - \bm{v}$.
By checking the KKT conditions, we obtain the optimal solution to \eqref{ADMMStep1Solving} is
\begin{align}
\begin{pmatrix}
x_k\\
z_k
\end{pmatrix}
&= \left\{
\begin{aligned}
&( x_k(0), z_k(0) )^T,&& \text{if }g_k( x_k(0), z_k(0) ) \leq 0,\\
&( x_k(\mu_k),z_k(\mu_k))^T,&&\text{else},
\end{aligned}\right. \nonumber \\
y_n &= \left\{
\begin{aligned}
&\tau_{3,n},&& \text{if }|\tau_{3,n}| \leq  1,\\
&\tau_{3,n}/|\tau_{3,n}|,&&\text{else},
\end{aligned}\right.  \label{SolutionMulticast1}
\end{align}
where $1\leq k\leq K$, $1\leq n\leq N$, $x_k(\mu) \triangleq \mu  + \tau_{1,k}$, $z_k(\mu) \triangleq - \frac{1}{2}\mu  + \tau_{2,k}$, and $\mu_k =
\frac{2}{5}\left( \tau_{2,k} - 2\Re\{\tau_{1,k}\} - s_k \right)$.

\emph{Solution to \eqref{ADMMStep2}:} Based on \eqref{ALForMultiCast}, \eqref{ADMMStep2} becomes the following unconstraint convex quadratic problem
\begin{align}
\min_{\gamma,\bm{\phi}}&~\frac{\rho}{2}\left\|\bm{\tau}_4 - \bm{T}^H\bm{\phi}\right\|^2
+ \frac{\rho}{2}\left\|\bm{\tau}_5 - \bm{\phi}\right\|^2
+ \frac{\rho}{2}\left\|\bm{\tau}_6 -  \gamma\bm{1}\right\|^2
- \gamma, \nonumber
\end{align}
where $\bm{\tau}_4 \triangleq \bm{x} + \bm{u}$, $\bm{\tau}_5 \triangleq \bm{y} + \bm{v}$, and $\bm{\tau}_6 \triangleq \bm{z} + \bm{w}$. By letting the first order derivative to be zero, and the optimal solution to \eqref{ADMMStep2} is given by
\begin{align}
\gamma = \frac{1 + \rho\bm{1}^T\bm{\tau}_{6}}{\rho K},~ \bm{\phi} = (\bm{I} + \bm{T}\bm{T}^H)^{-1}( \bm{\tau}_5 + \bm{T}\bm{\tau}_4 ).
\label{SolutionMulticast2}
\end{align}

In summary, in this section, we have presented a method to solve \eqref{FirstForm}.
The proposed method repeatedly solving \eqref{MultiCastSCAed} by using the closed-form iterations in \eqref{ADMMStep}. We note that for different value of $k$ ($n$), updating $\{x_k,z_k\}$ ($\{y_n\}$) can be implemented in a parallel manner which potentially reduces the time consumption for computation. Besides, by carefully checking the iteration process in \eqref{ADMMStep}, the computational complexity (evaluated through the number of float multiplications) is $J_1\times \left( O(N^3) + J_2\times (O(N^2) +  O(K)) \right)$, where $J_2$ is the number of the iterations that are required for \eqref{ADMMStep} to converge and $J_1$ is the number of times for which we repeatedly solve \eqref{MultiCastSCAed}. Note that the computation complexity of the SDR-based method in Section III-A is $O(N^6)$.

\section{Multi-user downlink transmission}
In this section, we design passive beamforming for the case where the BS sends individual messages for different users.
For $\forall k,k^{\prime}$, define $\bm{\alpha}_{k,k^{\prime}} \triangleq \frac{1}{\sigma_k}\bm{H}_k^H\bm{G}\bm{b}_{k^{\prime}}$, $\beta_{k,k^{\prime}} \triangleq \frac{1}{\sigma_k} \bm{b}_{k^{\prime}}^H\bm{f}_k$, $ \bm{\Lambda}_k \triangleq [ \bm{\alpha}_{k,1}, \cdots, \bm{\alpha}_{k,k-1}, \bm{\alpha}_{k,k+1}, \cdots, \bm{\alpha}_{k,K} ]$, and
$\hat{\bm{\beta}}_k \triangleq [\beta_{k,1}, \cdots, \beta_{k,k-1}, \beta_{k,k+1}, \cdots, \beta_{k,K}]^T$.
Define $d_k(\bm{\phi},\gamma)
\triangleq \frac{|\bm{\alpha}_{k,k}^H\bm{\phi} + \beta_{k,k}|^2}{\gamma}$.
Based on \eqref{SNRbroadcast}, \eqref{FirstForm} can be rewritten as
\begin{subequations}
\label{BroadCastSCAed}
\begin{align}
&\max_{\bm{\phi},\gamma}\quad \gamma,\\
&~{\rm s.t.}~
1 + ||\bm{\Lambda}_k^H \bm{\phi}  + \hat{\bm{\beta}}_k||^2-d_k(\bm{\phi},\gamma)\leq 0, \forall k,\label{broadCastNonConvex}\\
&~~~~~~  |\phi_n|^2 \leq 1,~\forall n
\end{align}
\end{subequations}
%where for $\forall k,k^{\prime}$, $\bm{\alpha}_{k,k^{\prime}} \triangleq \frac{1}{\sigma_k}\bm{H}_k^H\bm{G}\bm{b}_{k^{\prime}}$, $\beta_{k,k^{\prime}} \triangleq \frac{1}{\sigma_k} \bm{b}_{k^{\prime}}^H\bm{f}_k$, $ \bm{\Lambda}_k \triangleq [ \bm{\alpha}_{k,1}, \cdots, \bm{\alpha}_{k,k-1}, \bm{\alpha}_{k,k+1}, \cdots, \bm{\alpha}_{k,K} ]$,
%$\hat{\bm{\beta}}_k \triangleq [\beta_{k,1}, \cdots, \beta_{k,k-1}, \beta_{k,k+1}, \cdots, \beta_{k,K}]^T$,
%and  $q(\bm{\phi},\gamma)
%\triangleq \frac{|\bm{\alpha}_{k,k}^H\bm{\phi} + \beta_{k,k}|^2}{\gamma}$.

The difficulty of solving \eqref{BroadCastSCAed} lies in the fact that $-d_k(\bm{\phi},\gamma)$ in \eqref{broadCastNonConvex} is non-convex. In fact, $d_k(\bm{\phi},\gamma)$ is a convex quadratic-over-linear function, meaning that
$-d_k(\bm{\phi},\gamma)$ is a concave function of $(\bm{\phi},\gamma)$.
Following the principle of SCA \cite{M.Razaviyayn}, we replace $-d_k(\bm{\phi},\gamma)$ with its first order Taylor expansion at some feasible point $(\bm{\phi}_e,\gamma_e)$, which constitutes an upper bound on $-d_k(\bm{\phi},\gamma)$,  and obtain the following convex constraint,
\begin{align}
2\Re \{\hat{\bm{t}}_k^H \bm{\phi}\} + \hat{s}_k \geq q_k\gamma
+ ||\bm{\Lambda}_k^H \bm{\phi}  + \hat{\bm{\beta}}_k||^2,\forall k,\label{BroadCastSNRConstraint}
\end{align}
where $\hat{\bm{t}}_k \triangleq  \frac{1}{\gamma_e}( \bm{\alpha}_{k,k}^H\bm{\phi}_e + \beta_{k,k} )\bm{\alpha}_{k,k}$,
$\hat{s}_k \triangleq 2 \frac{|\bm{\alpha}_{k,k}^H\bm{\phi}_e + \beta_{k,k}|^2}{\gamma_e} - \frac{2\Re \{(\bm{\phi}_e^H\bm{\alpha}_{k,k} + \beta_{k,k}^*)\bm{\alpha}_{k,k}^H\bm{\phi}_e\}}{\gamma_e} - 1$, and
$q_k \triangleq \frac{|\bm{\alpha}_{k,k}^H\bm{\phi}_e + \beta_{k,k}|^2}{\gamma_e^2}$.

Based on \eqref{BroadCastSNRConstraint}, we obtain the convex problem below
\begin{align}
\label{ADMMForBroadCast}
\max_{\hat{\mathcal{X}}} ~\gamma,
~~{\rm s.t.}~\left\{\begin{aligned}
&\hat{g}(\bm{x}_k,z_k) \geq 0,~\forall k, \\
&|y_n|^2 \leq 1,~\forall n,~\bm{x}_k = \bm{T}^H_k\bm{\phi}, ~\forall k, \\
&\bm{y} = \bm{\phi},\quad \bm{z} = \gamma\bm{1},
\end{aligned}\right.
\end{align}
where $\hat{\mathcal{X}} = \{\{\bm{x}_k\}_{k=1}^K, \bm{y},\bm{z},\bm{\phi},\gamma\} $, $\hat{g}(\bm{x}_k,z_k) \triangleq 2\Re \{x_{k,1}\} + \hat{s}_k - q_k z_k - ||\bm{x}_{k,\bar{1}}  + \hat{\bm{\beta}}_k||^2$, and $\bm{T}_k \triangleq [\hat{\bm{t}}_k, \bm{\Lambda}_k]$ for $\forall k$.
Note that in \eqref{ADMMForBroadCast}, we have introduced slack variables $\{ \{\bm{x}_k\}_{k=1}^K, \bm{y}, \bm{z}\}$, which is similar to what we have done in \eqref{ADMMForMultiCast}.
%It is worth noting that the main difference between \eqref{ADMMForBroadCast} and \eqref{ADMMForMultiCast} is that in order to decompose the $K$ constraints in \eqref{BroadCastSNRConstraint}, $K$ vector-valued variables, i.e., $\{\bm{x}_k\}_{k=1}^K$, are required due to the quadratic term in the right-hand-side of \eqref{BroadCastSNRConstraint}, which comes from the mutual interference among different users, while for the multicast case in \eqref{ADMMForMultiCast}, $K$ scaler variables, i.e., $(x_1,x_2,\cdots,x_K)$, are enough to decompose the $K$ SNR constraints.

Based on \eqref{ADMMForBroadCast}, the augmented Lagrangian function is
\begin{align}
\mathcal{L}_{\rho}^{(\mathrm{b})} = &-\gamma + \frac{\rho}{2}\sum_{k=1}^K ||\bm{x}_k - \bm{T}_k^H \bm{\phi} + \bm{u}_k|| \nonumber \\
&\quad +\frac{\rho}{2} ||\bm{y} - \bm{\phi} + \bm{v}||^2  +\frac{\rho}{2} ||\bm{z} - \gamma\bm{1} + \bm{w}||^2.
\label{LarBroad}
\end{align}
where $\bm{u}_k$ is the dual variable corresponding to the constraint $\bm{x}_k = \bm{T}_k^H\bm{\phi}$ for $1\leq k\leq K$.
By dividing $\hat{\mathcal{X}}$ into two groups, i.e., $\hat{\mathcal{X}}_1 = \{\{\bm{x}_k\}_{k=1}^K,\bm{y},\bm{z}\}$ and $\hat{\mathcal{X}}_2 = \{\bm{\phi},\gamma\}$, we now use ADMM algorithm solve \eqref{ADMMForBroadCast}.
%by iteratively updating $\hat{\mathcal{X}}_1$, $\hat{\mathcal{X}}_2$, and $\{\{\bm{u}_k\}_{k=1}^K,\bm{v},\bm{w}\}$ in a similar way to \eqref{ADMMStep}. in a similar way to \eqref{ADMMStep}.
In fact, to solve \eqref{ADMMForBroadCast}, the ADMM iteration process is essentially the same as \eqref{ADMMStep}, and the differences appear only when we update  $\{\bm{x}_k,z_k\}$ and $\bm{\phi}$, which we present below.

\emph{The update of $\{\bm{x}_k,z_k\}$ for $k = 1,2,\cdots,K$:} according to \eqref{LarBroad}, the  update of $\{\bm{x}_k,z_k\}$ involves solving the following optimization problem,
\begin{align}
\label{broadcastADMMStepxz}
\min_{\bm{x}_k,z_k}~||\bm{x}_k - \bm{\tau}_1^{(k)} || + (z_k - \tau_{2,k})^2,~{\rm s.t.}&~\hat{g}(\bm{x}_k,z_k) \geq 0,
\end{align}
where $\bm{\tau}_1^{(k)} \triangleq \bm{T}_k^H \bm{\phi} - \bm{u}_k$. By checking the KKT condition of \eqref{broadcastADMMStepxz}, we obtain the optimal solution of \eqref{broadcastADMMStepxz} as
\begin{align}
(\bm{x}_k,z_k) = \left\{
\begin{aligned}
&( \bm{x}_k(0), z_k(0) ),&& \text{if }\hat{g}_k( \bm{x}_k(0), z_k(0) ) \leq 0 \\
&( \bm{x}_k(\mu_k),z_k(\mu_k)),&& \text{else }
\end{aligned}
\right.\nonumber
\end{align}
where $x_{k,1}(\mu) = \tau_{1,1}^{(k)} + \mu$, $\bm{x}_{k,\bar{1}}(\mu) = (1 + \mu)^{-1}( \bm{\tau}_{1,\bar{1}}^{(k)} - \mu \hat{\bm{\beta}}_k)$,
$z_k(\mu) = -\frac{\mu q_k}{2} + \tau_{2,k}$,
$\bm{\tau}_1^{(k)} = \bm{T}_k^H \bm{\phi} - \bm{u}_k$, and $\mu_k$ is the root of equation $\hat{g}_k(\bm{x}_{k}(\mu),z_k(\mu)) = 0$ in $(0,\infty)$. Note that based on the expressions of  $\bm{x}_k(\mu)$ and $z_k(\mu)$, $\hat{g}_k(\bm{x}_{k}(\mu),z_k(\mu)) = 0$ is actually a cubic equation with respect to $\mu$. Therefore $\mu_k$ can be written in a closed form, which we omit due to the space limitation.

\emph{The update of $\bm{\phi}$}:
Based on \eqref{LarBroad}, the update of $\bm{\phi}$ is given by
$
\bm{\phi} = \mathrm{argmin}_{\bm{\phi}}~ \left\{\sum_{k=1}^K||\bm{\tau}_4^{(k)} - \bm{T}_k^H \bm{\phi} || + ||\bm{\tau}_5 - \bm{\phi} ||^2\right\}
$
where $\bm{\tau}_4^{(k)} \triangleq \bm{x}_k + \bm{u}_k$ for $k = 1,\cdots,K$.
Note that this is a unconstraint quadratic problem, and the solution is
\begin{align}
\bm{\phi} = \left(\bm{I} + \sum_{k=1}^K\bm{T}_k\bm{T}_k^H\right)^{-1}\left( \bm{\tau}_5 +  \sum_{k=1}^K \bm{T}_k\bm{\tau}_4^{(k)} \right).
\end{align}

In summary, in this section, we have proposed a method to solve \eqref{FirstForm} for the multi-user downlink transmission scenario.
The computational complexity of the method in this section is higher than that in the Section III due to the fact that more slack variables are introduced. However, the update of $\{\bm{x}_k,z_k\}_{k=1}^K$ and $\{y_n\}_{n=1}^N$ can be implemented in parallel manner, which is helpful to reduce the computing time.

\section{Numeric Result}
In this section, numeric results are presented to show the performance of the IRS-aided wireless networks.
Unless specified, we set $\sigma_1^2 = \cdots=\sigma_K^2 = -40$ dBm, the transmit power of the BS as $P_B = 10$ dBm, $M = 30$, and $K = 15$. The path-loss exponents are set to be $3$ for the channels from the BS to the IRS and to the users, and are set to be $2$ for the channels from the IRS to the users. The locations of the BS and the IRS are $(-50,0)$ and $(0,30)$, respectively. The locations of the users are uniformly and randomly generated within $(-20,20)\times(-20,20)$.

%\begin{table}[t]
%\small
%\centering
%\begin{tabular}{|c|c|c|}
%\hline\hline
%\multirow{2}{*}{\begin{tabular}[c]{@{}c@{}} SDR-Based \\Method\end{tabular}}
%&ACT& 5.9464 sec \\
%\cline{2-3} & $\min_k~{\rm SNR}_k$ & 4.2310\\
%\hline
%\multirow{2}{*}{\begin{tabular}[c]{@{}c@{}} ADMM-Based \\Method \end{tabular}}
%&ACT& 0.5881 sec\\
%\cline{2-3} &$\min_k~{\rm SNR}_k $ & 4.6951 \\
%\hline
%\hline
%\end{tabular}
%\caption{Performance comparisons between the SDR-based method and the proposed ADMM-based method with $N = 200$.}\label{MethodComparision}
%\vspace{-5mm}
%\end{table}

\begin{table}[t]
\caption{Comparisons between the SDR-based method and the proposed ADMM-based method with $N = 200$.}\label{MethodComparision}
\centering
\begin{tabular}{|@{}c@{}|c|c|}
\hline\hline
\diagbox[width=9.5em, height=1.3em]{ }{ }& ACT (secs)& $\min_k~{\rm SNR}_k$ \\
\hline
\begin{tabular}[c]{@{}c@{}} SDR, Scheme 1 \end{tabular}
&  5.21   & 4.9239\\
\hline
\begin{tabular}[c]{@{}c@{}} SDR, Scheme 2\end{tabular}
&  5.21   & 3.2342 \\
\hline
\begin{tabular}[c]{@{}c@{}} ADMM, subject to \eqref{IdealRS} \end{tabular}
&  0.49    & 5.4905  \\
\hline
\begin{tabular}[c]{@{}c@{}} ADMM, subject to \eqref{NotIdealIRS} \end{tabular}
&  0.49    & 5.3901 \\
\hline
\hline
\end{tabular}
\vspace{-5mm}
\end{table}
To demonstrate the computational efficiency of the proposed ADMM-based method, we compare it with the benchmark SDR-based method introduced in Section III-A in terms of: 1) the average CPU time (ACT), and 2) the obtained communication performance, i.e., $\min_k~{\rm SNR}_k$.
We run both methods using software MATLAB R2016b.
For the ADMM-based method, we set the maximum iteration number of \eqref{ADMMStep} as $2000$, and we repeatedly solve \eqref{MultiCastSCAed} using the iteration in \eqref{ADMMStep} for $5$ times.
For the SDR-based method, we use CVX with solver SDPT3 for implementation \cite{M.GrantCVX}, and
if the solution is not rank-one,
we consider the following two schemes to recover a feasible solution.
For notational simplicity, we use $\bm{\Psi}$ to denote the optimal solution to problem \eqref{SDPBenchmark} and $\bm{\phi} = [\phi_1,\cdots,\phi_n]$ to denote the recovered solution.

\textbf{Scheme 1:} Denote $\tilde{\bm{\phi}}\in \mathcal{C}^{(N+1)\times 1}$ as a randomly generated vector using Gaussian distribution $\mathcal{CN}(\bm{0},\bm{\Psi})$. Let $\hat{\bm{\phi}}\in \mathcal{C}^{N \times 1}$ with $\hat{\phi}_n = \tilde{\phi}_n/\tilde{\phi}_{N+1}$ for $1\leq n\leq N$. Then, we set $\phi_n = \hat{\phi}_n/|\hat{\phi}_n|$ for $1\leq n\leq N$.
Note this scheme is also used in existing works, see e.g., \cite{Q.WuTWC2019,Q.WuGLOBLE2018}.

\textbf{Scheme 2:} Denote $\tilde{\bm{\phi}}\in \mathcal{C}^{(N+1)\times 1}$ as the eigenvector of $\bm{\Psi}$ that corresponds to the largest eigenvalue. Let $\hat{\bm{\phi}}\in \mathcal{C}^{N \times 1}$ with $\hat{\phi}_n = \tilde{\phi}_n/\tilde{\phi}_{N+1}$ for $1\leq n\leq N$. Then, we set $\phi_n = \hat{\phi}_n/|\hat{\phi}_n|$ for $1\leq n\leq N$.

Note that the solutions generated by Scheme 1 and Scheme 2 are feasible to both constraints \eqref{IdealRS} and \eqref{NotIdealIRS}.
In our simulation, if Scheme 1 is used, we randomly generate $10^4$ solutions and pick the best one.
For the proposed ADMM-based method, if constraint \eqref{IdealRS} is considered, we directly compute the solution by using the methods presented in the previous sections. If constraint \eqref{NotIdealIRS} is considered, we directly normalize all the RCs obtained by the using proposed ADMM-based method so that the solution is feasible to \eqref{NotIdealIRS}.
The ACTs are obtained by using the timing instructions of MATLAB, i.e., 'tic' and 'toc', and are averaged across 50 random channel realizations.
We summarize the comparison results in Table \ref{MethodComparision}.
Note the if Scheme 1 is used, the CPU time for the randomization process is not taken into account, and thus the results in Table \ref{MethodComparision}, in fact, underestimate the time consumed by the SDR-based method with Scheme 1.
From Table \ref{MethodComparision}, we can see that the proposed ADMM-based method runs much faster and achieves better performance than the benchmark SDR-based method.

\begin{figure}
  \centering
  \includegraphics[width=3.5in]{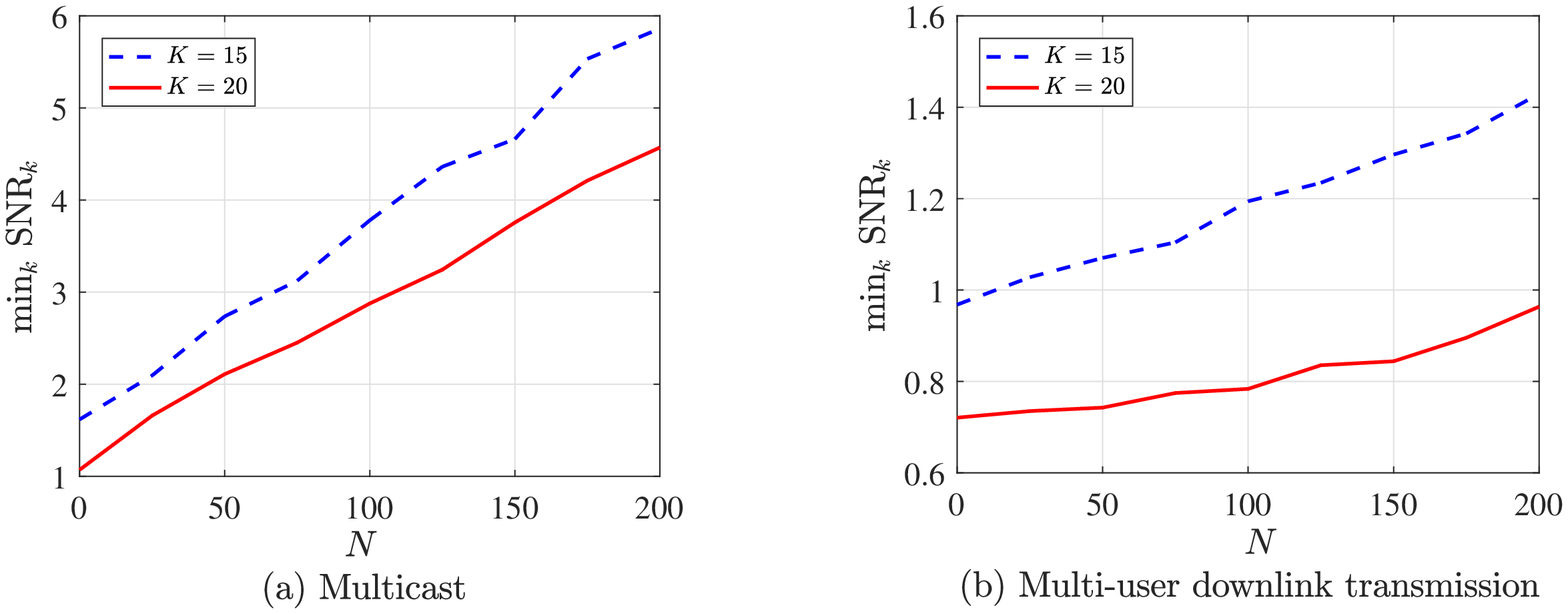}\\
  \caption{${\rm SNR}$ balancing level versus the number of reflecting elements $N$.}\label{FigOne}
  \vspace{-5mm}
\end{figure}

In Fig. \ref{FigOne}, we plot the smallest SNR of the users versus the number of reflecting elements, $N$.
In our simulation, for the case of multicasting, we use the method in \cite{N.D.SidiropoulosTSP2006} to optimize the beamforming vector of the BS, i.e., $\bm{b}$, and for the multi-user downlink transmission scenario, we use the method in \cite{M.SchubertTVT2004} to optimize the beamforming matrix of the BS, i.e., $[\bm{b}_1,\bm{b}_2,\cdots,\bm{b}_K]$. The results in Fig. \ref{FigOne} are obtained by alternatingly optimizing the active beamforming at the BS and the passive beamforming at the IRS, and are averaged over $50$ random channel realizations. Note that in Fig. \ref{FigOne},  $N = 0$ stands for the case where there is no IRS in the system.
Based on the results in Fig. \ref{FigOne}, we can see that due to the deployment of the IRS, SNRs of the users can be significantly improved, especially for the scenario of multicasting.

\begin{figure}
  \centering
  \includegraphics[width=3.5in]{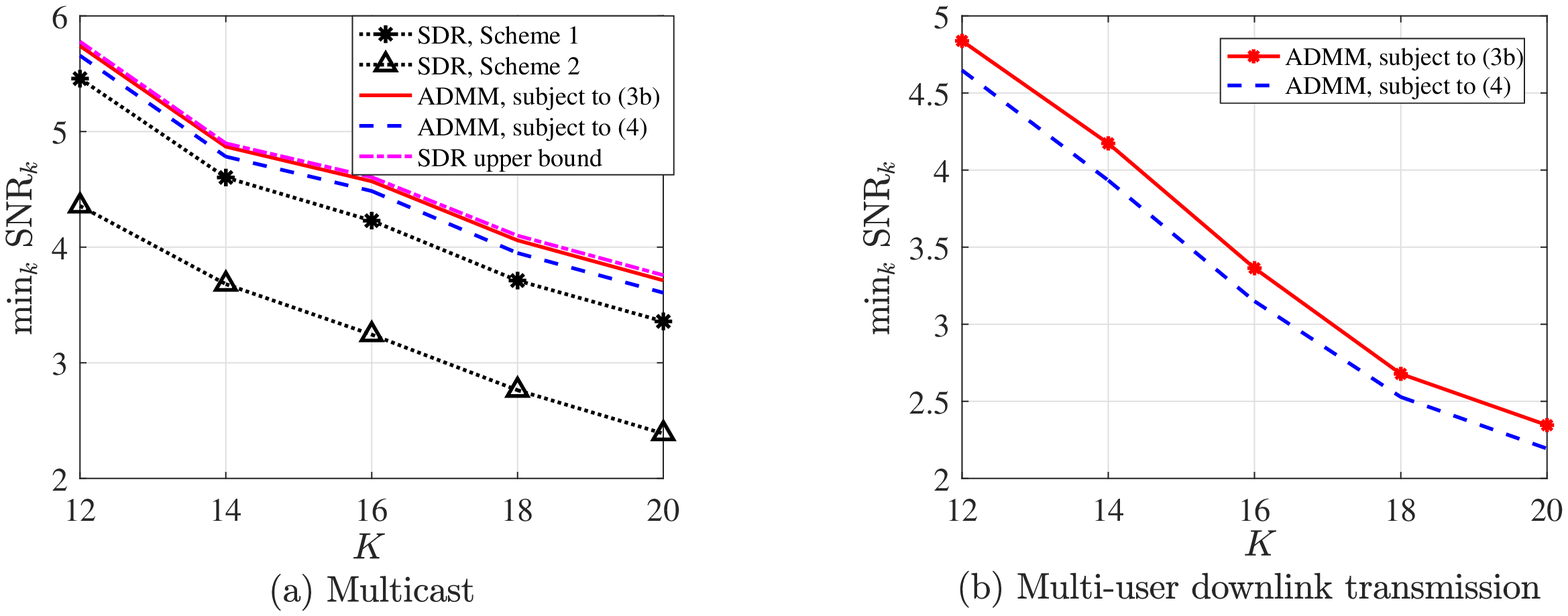}\\
  \caption{${\rm SNR}$ balancing level versus the number of users $K$, where we set $N = 150$.}\label{FigTwo}
  \vspace{-5mm}
\end{figure}

In Fig. \ref{FigTwo}, we plot the smallest SNR of the users versus the number of the users $K$, where we set $P_B = 10$ dBm in Fig. \ref{FigTwo}(a) and $P_B = 15$ dBm in Fig. \ref{FigTwo}(b).
In Fig. \ref{FigTwo}(a), the optimal value of problem \eqref{SDPBenchmark} is plotted, which constitutes an upper bound on the optimal value of \eqref{FirstForm} due to the neglected rank-one constraint.
First of all, Fig. \ref{FigTwo}(a) reveals that the performance of the ADMM-based method approaches the upper bound, which indicates that the solution obtained by the ADMM-based method is nearly optimal.
Besides, from Fig. \ref{FigTwo}, we can see that by using the ADMM-based method, the performance loss is relatively small when the RCs are subject to \eqref{NotIdealIRS}.
In fact, through extensive numeric experiments, we find that using the ADMM-based method, the amplitudes of most RCs are equal to one after optimization. Note that the non-convex constraint \eqref{NotIdealIRS} is generally hard to handle, and Fig. \ref{FigTwo} inspires us that if $\bm{\phi}$ is subject to \eqref{NotIdealIRS}, we can replace \eqref{NotIdealIRS} with \eqref{IdealRS} to optimize $\bm{\phi}$ and
recover a feasible solution by a simple projection operation, which does not cause much performance loss.
In Fig. \ref{FigTwo}(a), we can also see that for the SDR-based method, Scheme 1 achieves better performance than Scheme 2.
This is because given that $\bm{\Phi}$ is not rank-one, the solution generated by Scheme 2 is sub-optimal, but in Scheme 1, due to the fact that multiple solutions are randomly generated, it is possible that a few of the solutions are close to the optimal solution, which leads to a good performance.
It is important to note that in Fig. \ref{FigTwo}(a), the proposed ADMM-based method exhibits better performance than the SDR-based method no matter Scheme 1 or Scheme 2 is used to recover a feasible solution, which demonstrates the superiority of the ADMM-based method.

\section{Conclusions}
We investigated the passive beamforming design problem in IRS-aided multi-user systems.
We maximized the smallest SNR of the multiple users. Low complexity methods based on ADMM algorithm were proposed to solve the established optimization problems. Simulation results demonstrated the efficiency of the proposed ADMM-based method.


\begin{thebibliography}{99}
\bibitem{C.Liaskos2018}
C. Liaskos, S. Nie, A. Tsioliaridou, A. Pitsillides, S. Ioannidis, and I. Akyildiz, ``A new wireless communication paradigm through software-controlled metasurfaces,'' \emph{IEEE Commun. Mag.}, vol. 56, no. 9, pp. 162--169, Sep. 2018.

\bibitem{Q.WuCMCM2020}
Q. Wu and R. Zhang, ``Towards smart and reconfigurable environment: Intelligent reflecting surface aided wireless network,'' \emph{IEEE Commun. Mag.}, vol. 58, no. 1, pp. 106--112, Jan. 2020.

\bibitem{Q.WuTWC2019}
Q. Wu and R. Zhang, ``Intelligent reflecting surface enhanced wireless network via joint active and passive beamforming,'' \emph{IEEE Trans. Wireless Commun.}, vol. 18, no. 11, pp. 5394--5409, Nov. 2019.

%\bibitem{Q.WuICASSP2019}
%Q. Wu and R. Zhang, ``Beamforming optimization for intelligent reflecting surface with discrete phase shifts,'' \emph{IEEE International Conference on Acoustics, Speech and Signal Processing (ICASSP)}, Brighton, UK, 2019, pp. 7830--7833.

\bibitem{C.HuangTWC2019}
C. Huang, A. Zappone, G. C. Alexandropoulos, M. Debbah, and C. Yuen, ``Reconfigurable intelligent surfaces for energy efficiency in wireless communication,'' \emph{IEEE Trans. Wireless Commun.}, vol. 18, no. 8, pp. 4157--4170, Aug. 2019.

\bibitem{Y.ZhengWCL}
Y. Zheng, S. Bi, Y. J. Zhang, Z. Quan, and H. Wang, ``Intelligent reflecting surface enhanced user cooperation in wireless powered communication networks,'' \emph{IEEE Wireless Commun. Let.} accepted to appear.



\bibitem{X.GuanCL2020}
X. Guan, Q. Wu and R. Zhang, ``Joint power control and passive beamforming in IRS-assisted spectrum sharing,'' \emph{IEEE Commun. Let.} accepted to appear.

%\bibitem{Z.Chu2020WCL}
%Z. Chu, W. Hao, P. Xiao, and J. Shi, ``Intelligent reflecting surface aided multi-antenna secure transmission,'' \emph{IEEE Wireless Commun. Let.}, vol. 9, no. 1, pp. 108--112, Jan. 2020.
%
%\bibitem{H.ShenCL2019}
%H. Shen, W. Xu, S. Gong, Z. He, and C. Zhao, ``Secrecy rate maximization for intelligent reflecting surface assisted multi-antenna communications,'' \emph{IEEE Commun. Let.}, vol. 23, no. 9, pp. 1488--1492, Sep. 2019.
\bibitem{LimengDongFx}
L. Dong and H.-M. Wang, ``Secure MIMO transmission via intelligent reflecting surface,'' \emph{IEEE Wireless Commun. Let.}, vol. 9, no. 6, pp. 787-790, Jun. 2020.


\bibitem{S.boydADMM2011}
S. Boyd, N. Parikh, E. Chu, B. Peleato, and J. Eckstein, ``Distributed optimization and statistical learning via the alternating direction method of multipliers,'' \emph{Found. Trends Mach. Learn.}, vol. 3, no. 1, pp. 1--122, 2011.


\bibitem{N.D.SidiropoulosTSP2006}
N. D. Sidiropoulos, T. N. Davidson, and Zhi-Quan Luo, ``Transmit beamforming for physical-layer multicasting,'' \emph{IEEE Trans. Signal Process.}, vol. 54, no. 6, pp. 2239--2251, Jun. 2006.

\bibitem{M.SchubertTVT2004}
M. Schubert and H. Boche, ``Solution of the multiuser downlink beamforming problem with individual SINR constraints,'' \emph{IEEE Trans. Veh. Tech.}, vol. 53, no. 1, pp. 18--28, Jan. 2004.

\bibitem{Q.WuGLOBLE2018}
Q. Wu and R. Zhang, ``Intelligent reflecting surface enhanced wireless network: Joint active and passive beamforming design,'' \emph{2018 IEEE Global Communications Conference (GLOBECOM)}, Abu Dhabi, United Arab Emirates, 2018.

\bibitem{A.WieselTSP}
A. Wiesel, Y. C. Eldar, and S. Shamai,``Linear precoding via conic optimization for fixed MIMO receivers,'' \emph{IEEE Trans. Signal Process.}, vol. 54, no. 1, pp. 161--176, Jan. 2006.


\bibitem{M.Razaviyayn}
M. Razaviyayn, ``Successive convex approximation: Analysis and applications,'' Ph.D. dissertation, Univ. Minnesota, Minneapolis, MN, USA,
2014.

\bibitem{M.GrantCVX}
M. Grant and S. Boyd. (2016). \emph{CVX: MATLAB Software for Disciplined
Convex Programming.} [Online]. Available: http://cvxr.com/cvx.

\end{thebibliography}
\end{document}